\documentclass[aps,prd,twocolumn,groupaddress]{revtex4}
\usepackage{graphicx} 
\usepackage{color}
\usepackage{xcolor}
\usepackage[colorlinks=true, allcolors=black]{hyperref}
\usepackage{xspace}

\def\km3net{KM3NeT}
\newcommand{\neutrino}{KM3-230213A\xspace}
\begin{document}

\title{KM3-230213A:\\ An Ultra-High Energy Neutrino from a Year-Long Astrophysical Transient}
\author{Andrii Neronov$^{1,2}$, Foteini Oikonomou$^{3}$, Dmitri Semikoz$^1$}

\affiliation{$^{1}$Universit\'e Paris Cit\'e, CNRS, Astroparticule et Cosmologie, 
F-75013 Paris, France}

\affiliation{$^{2}$Laboratory of Astrophysics, \'Ecole Polytechnique F\'ed\'erale de Lausanne, CH-1015 Lausanne, Switzerland}

\affiliation{$^3$Institutt for fysikk, NTNU, Trondheim, Norway}

\begin{abstract}
    The \km3net\ collaboration has recently reported the detection of a neutrino event with energy in excess of 100 PeV. This detection is in 2.5-3$\sigma$ tension with the upper limit on the neutrino flux at this energy imposed by IceCube and the Pierre Auger Observatory, if the event is considered part of the diffuse all-sky neutrino flux. We explore an alternative possibility that the event originates from a flare of an isolated source. We show that the data of \km3net, IceCube and the Pierre Auger Observatory are consistent with the possibility of a source flare of duration $T\lesssim 2$~yr with muon neutrino flux $F\sim 3\times 10^{-10}(1\mbox{ yr}/T)$~erg/(cm$^2$s). Constraints on the neutrino spectrum indicate that the protons responsible for the neutrino emission have a very hard spectrum in the $E_p\gtrsim 10^{19}$~eV energy range, or otherwise that the neutrinos are produced by photohadronic interactions with infrared photons. The all-sky rate of similar neutrino flaring sources is constrained to be ${\cal R}\lesssim 0.4/$yr. 
\end{abstract}
\maketitle

\section{Introduction}

The \km3net\ Collaboration has recently reported the detection of the neutrino \neutrino with energy 110-790 PeV~\cite{KM3NeT2025}. This event was reported to be in 2.5$\sigma$-3$\sigma$ tension with the non-observation of neutrinos with similar energy by the Pierre Auger Observatory~\cite{PierreAuger:2019azx,PierreAuger:2023pjg} and IceCube~\cite{IceCube:2025ezc,KM3NeT:2025global_view}.  In accompanying papers the possible blazar origin~\cite{KM3NeT:2025bxl}, or Galactic origin~\cite{KM3NeT:2025galactic} were investigated. It was concluded that the origin of the neutrino is inconsistent with a Galactic source. Even though no firm conclusion on the possible extragalactic source origin of the event was made, it remains unclear why no neutrinos of such high energy have been seen with IceCube or the Pierre Auger Observatory in association with any type of bright extragalactic astrophysical sources. 

In what follows we show that the \km3net\ signal is in fact inconsistent also with the extragalactic source hypothesis, unless the source is assumed to undergo an outburst during the \km3net\ exposure or have a transient nature altogether. We discuss constraints on the parameters of the outburst that is responsible for the observed neutrino event and also on the population of similar bursting sources in the Universe.

\section{An isolated source model for the neutrino event}

We consider an isolated source that produces a flare with the flux level $F$ over a time interval $T$, which we assume to be much larger than one day (so that the time-averaged instrument response functions can be used in the estimates below). Within this model, a steady source is one for which the time scale $T$ is comparable to or larger than the overall IceCube and Pierre Auger Observatory exposures (decade-scale). 

For simplicity, we assume that neutrino events have energy close to the estimated energy of the ARCA event, $E_\nu\sim 220$~PeV. The number of neutrino events from the flare in ARCA and IceCube detectors is 
\begin{eqnarray}
    &&N_{\rm ARCA}=E_\nu^{-1}F\cdot T\cdot A_{\rm ARCA}\nonumber\\
    &&N_{\rm IC}=E_\nu^{-1}F\cdot T\cdot A_{\rm IC}\nonumber
\end{eqnarray}
where $A_{\rm ARCA}, A_{\rm IC}$ are the time-averaged effective areas of the ARCA and IceCube detector in the direction of the source. 
This implies the relation between the event statistics in IceCube and ARCA,
\begin{equation}
    N_{\rm IC}=N_{\rm ARCA}\frac{A_{\rm IC}}{A_{\rm ARCA}}.
\end{equation}

The effective area of IceCube in the source direction can be extracted from the public IceCube data \cite{IceCube:2021xar}, $A_{\rm IC}\simeq 6\times 10^3$~m$^2$. For ARCA, Ref.\ \cite{KM3NeT2025} only gives the sky-averaged effective area. We estimate the conversion from sky-average effective area to the effective area in the direction of the source for ARCA, using the information available for IceCube (both ARCA and IceCube are situated under a layer of material and should have comparable zenith angle dependence $A(Zd)$ of the effective area in the $100$~PeV energy range). Using this $A(Zd)$ template and considering the latitude of the ARCA detector, we find that the time-averaged area in the source direction is comparable to the sky-average area, so we adopt $A_{\rm ARCA}\simeq 4\times 10^2$~m$^2$ in the following calculations. 

The number of observed events in ARCA and IceCube are $N_{\rm ARCA,obs}=1$ and $N_{\rm IC,obs}=0$, during one year of observations by ARCA. The 90\% confidence interval of the ARCA measurement is 
\begin{equation}
0.05<N_{\rm ARCA}<4.7
\end{equation}
which can be converted into a prediction for the number of expected IceCube events:
\begin{equation}
1\left[\frac{T}{1\mbox{ yr}}\right]<N_{\rm IC,expected}<94\left[\frac{T}{1\mbox{ yr}}\right]
\end{equation}
The observation of zero events in IceCube suggests that at the 90\% confidence level, the number of IceCube events is 
$$
N_{\rm IC}< 2.3
$$
independently of the period of activity of the source $T$. Thus, if $T\lesssim 2$~yr, both the ARCA and IceCube event statistics are consistent (at 90\% confidence level) with a signal from one and the same source, with the ARCA event being an upward-fluctuation with respect to the average expected number of events within its one-year exposure and the IceCube non-detection being an under-fluctuation within with respect to the average expected number of events from the source activity episode. 

Note that the observed event statistics are clearly inconsistent with the possibility of a steady source with constant flux all over the time span of IceCube observations. The probability to detect zero events while the minimal possible expected event statistics is $\simeq 10$ is $p\simeq 5\times 10^{-5}$.

\section{Properties of the flaring source}

Considering the $T\lesssim 2$~yr flaring source hypothesis, the model in which the ARCA event detection is consistent with the non-detection of neutrinos by IceCube, we can use the two datasets together to estimate the source flux. Combining the two Poisson probabilities (to have one or zero events for the expected signal statistics of $N_{\rm ARCA}$ events in ARCA and $20N_{\rm ARCA}$ events in IceCube) we can infer the most likely value of $N_{\rm ARCA}$ and a 90\% confidence interval on it,
$$
N_{\rm ARCA}=0.10_{-0.08}^{+0.18}.
$$
The muon neutrino flux of the source corresponding to this number is
\begin{equation}
F=\frac{E_\nu N_{\rm ARCA}}{A_{\rm ARCA}T}\simeq 2.9_{-2.3}^{+5.3}\times 10^{-10}\left[\frac{T}{1\mbox{ yr}}\right]^{-1}\frac{\mbox{erg}}{\mbox{cm}^2\mbox{s}}.
\label{eq:flux}
\end{equation}
The flux estimate is inversely proportional to the  duration of the flare, so that the total fluence is fixed to
\begin{equation}
FT=\frac{E_\nu N_{\rm ARCA}}{A_{\rm ARCA}}\simeq 9_{-7}^{+16}\times 10^{-3}\frac{\mbox{erg}}{\mbox{cm}^2}.
\label{eq:fluence}
\end{equation}
The neutrino flux level of the flaring source is comparable to that of the electromagnetic power of the brightest Active Galactic Nuclei (AGN) and the fluence of the outburst is comparable to that of the brightest Gamma-Ray Bursts (GRBs). 

To produce such fluence, a source at a distance $D$ would need to release energy
\begin{equation}
    {\cal E}=2\pi\Theta_\nu^2 D^2(FT)
    \sim 10^{50}\left[\frac{D}{1\mbox{ Gpc}}\right]^2\left[\frac{\Theta_\nu}{1^\circ}\right]^2\mbox{ erg},
    \label{E_beam}
\end{equation}
\noindent where we have assumed a jet-like anisotropic emission pattern with jet opening angle $\Theta_\nu$. 
Such an energy output is consistent with AGN, GRBs and tidal disruption events. The most luminous GRBs ever detected have isotropic $\mathcal{E}_{\rm iso} \approx 10^{55}$~erg (see Table 5 of \cite{Burns:2023oxn}), but taking into account the spectral break and beamed emission detected by LHAASO corresponds to true energy output $\mathcal{E}=5.5 \cdot 10^{50} (\Theta/0.6^\circ)^2$ erg \cite{LHAASO:2023kyg}, which is comparable to Eq.(\ref{E_beam}).
The brightest known tidal disruption events have a somewhat lower energy output. For example, SWIFT J1644+57 had $L_{\rm iso}\approx 10^{48}\rm erg/s$ during the first weeks after detection and an isotropic energy output exceeding $\mathcal{E}_{\rm iso} = 10^{53}$~erg~\cite{Burrows:2011dn}.

\begin{figure*}
\includegraphics[width=1.95\columnwidth]{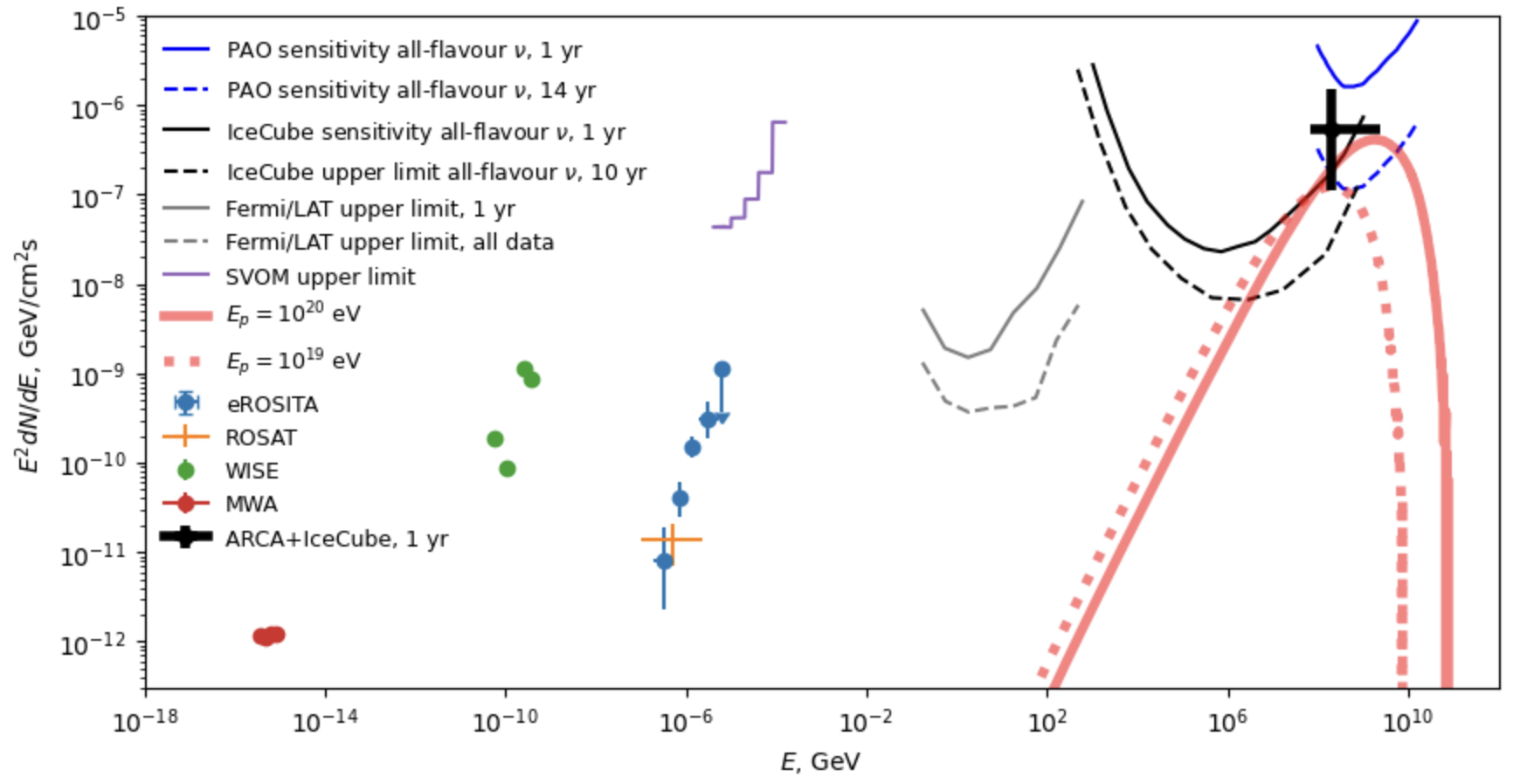}
    \caption{The black data point shows an estimate of the flux of a year-long neutrino flare from an isolated source, discussed in the text. This estimate is compared to the   spectral energy distribution of the source TXS 0614-083, the closest AGN to the ARCA event, extracted using Multi-Messenger Online Data Analysis service (MMODA) \url{https://mmoda.io}. Red data points are from the GLEAM radio survey of MWA \cite{Wayth:2015nla}. Green data points in the infrared are from WISE \cite{2010AJ....140.1868W}. Blue data points in X-rays are from eROSITA sky survey \cite{eROSITA:2024oyj}. The SVOM upper limit in the hard X-ray band is from Ref. \cite{KM3NeT:2025bxl}. The IceCube sensitivities are extracted using the SkyLLH software \cite{IceCube:2021mzg} available through the MMODA service. Fermi-LAT upper limits are extracted using the aperture photometry method, also implemented as an MMODA service. The blue neutrino upper limit gives the one-year (solid line) and 14 yr (dashed line) sensitivities of the Pierre Auger Observatory (PAO)~\cite{PierreAuger:2020llu}.}
    \label{fig:sed}
\end{figure*}

The difference in the effective areas of ARCA and IceCube leads to yet another constraint on the properties of the neutrino signal, as shown in Fig.\ \ref{fig:sed}. The IceCube sensitivity for a year-long exposure\footnote{Calculated using the Multi-Messenger Online Data Analysis (MMODA) service for IceCube, \url{https://mmoda.io}, using the instrument response functions from thepublic IceCube dataset \cite{IceCube:2021xar} and the SkyLLH analysis framework \cite{IceCube:2021mzg}.}, shown in this figure, is an envelope of the sensitivities for the  powerlaw-type spectra $dN_\nu/dE_\nu\propto E_\nu^{-\Gamma}$ for different slopes $\Gamma$. One can see that, for example, the upper limit on an $E_\nu^{-2}$ type power law (a horizontal line in $E^2\mbox{d}N/\mbox{d}E$ representation) would be much lower than the flux estimate obtained above. The upper limit would touch the envelope at the energy $E_\nu\sim 1$~PeV. This means that most of the statistics of the signal for this power-law slope would be expected at PeV energy, rather than in the 100~PeV range. Assuming that there was no observation of additional PeV neutrinos during the flare period, we find that the possibility of an $E^{-2}$ type neutrino spectrum is ruled out. 

The only possibility that appears consistent with the data is that the neutrino spectrum is hard, with $\Gamma\lesssim 1$. Remarkably, this is the slope of the low-energy tail of a neutrino spectrum that would be produced by a monoenergetic distribution of protons. As an example, the red solid and dashed curves in Fig.\ \ref{fig:sed} show the spectra of secondary neutrinos produced in interactions of $E_p=10^{19}$~eV and $E_p=10^{20}$~eV protons with low energy protons, computed using the AAfrag package \cite{Kachelriess:2022khq,Koldobskiy:2021nld}.

Another option to produce a neutrino spectrum with such a hard slope is through photohadronic interactions~\cite{1992A&A...253L..21M,MANNHEIM1995295,Atoyan:2001ey}. To avoid excess neutrinos at PeV energies, the target photons must have low enough energy, $E_{\gamma}\approx m_{\Delta}^2/40E_{\nu} \approx 0.2~\mbox{eV} (E_{\nu}/220~\mbox{PeV})^{-1}$, where we have assumed that the typical neutrino energy is 20 times lower than the energy of the parent proton, and $m_{\Delta} = 1232$~MeV is the mass of the $\Delta^+$ resonance \footnote{In the photohadronic scenario, the neutrino spectrum is strongly peaked due to the threshold energy for the process set by the mass of the $\Delta$-resonance, and the neutrino spectrum is described by a slope $\Gamma =\alpha + \beta -1$ where $\alpha$ and $\beta$ are the slopes of the high-energy protons and low-energy target photons respectively~\cite{Winter:2012xq}.}. Mid-infrared photons, such as those emitted by warm dust, are abundant in AGN tori. In blazars, infrared photons are the most natural—and potentially the only—abundant target photon field for interactions. Therefore, if \neutrino originated from a jetted AGN, its spectral shape and peak energy are consistent with photohadronic interactions in the infrared torus~\cite{Oikonomou:2021akf,podlesnyi2025}.

\section{Electromagnetic counterpart of the neutrino flare?}
\label{sec:electromagnetic}

Neutrino production in proton-proton or proton-photon interactions is always accompanied by the production of electromagnetic particles (gamma-rays, electrons, positrons) with the total energy in the electromagnetic channel being comparable to that of the neutrino channel. Gamma-rays produced together with neutrinos also initially have an energy distribution similar to that of the neutrinos. If gamma rays with energies in excess of 100~PeV escaped from an extragalactic source they would initiate an electromagnetic cascade in the interstellar medium of the source host galaxy and/or in the intergalactic medium. The electromagnetic cascade would redistribute the power initially injected in the highest energy gamma-rays to lower energy photons, perhaps down to the GeV-TeV band. The overall flux of such gamma-rays from the direction of the source may be much lower than the neutrino flux if the neutrino signal was initially emitted into a narrow beam with an opening angle $\Theta_\nu\ll 1$. The ratio of the neutrino and gamma-ray fluxes is expected to be 
\begin{equation}
    \left( \frac{F_\gamma}{F_\nu}\right)_{\Theta}=\frac{\Theta_\nu^2}{\Theta_{\gamma}^2}\sim 10^{-2}\left[\frac{\Theta_\nu}{1^\circ}\right]^2\left[\frac{\Theta_\gamma}{10^\circ}\right]^{-2}
\end{equation}
where $\Theta_{\gamma}$ is the opening angle of the electromagnetic jet. Another factor that leads to the suppression of the electromagnetic flux is the time delay of the cascade emission. If the distance scale of cascade delay is $D_\gamma$, the time delay is of the order of $T_\gamma\sim D_\gamma/c$ and an additional suppression factor due to this effect is 
\begin{equation}
     \left(\frac{F_\gamma}{F_\nu}\right)_{T}=\frac{T}{T_\gamma}\sim 10^{-3}\left[\frac{D_\gamma}{1\mbox{ kpc}}\right]^{-1}
    \left[\frac{T}{1\mbox{ yr}}\right]
\end{equation}
so that the electromagnetic flux associated with the neutrino flare might ultimately be difficult to detect. 

Alternatively, the electromagnetic cascade can develop right in the source. In this case, the time scale of cascade development can be short and no suppression of the electromagnetic flux due to the time delay of the cascade is present. This opens a possibility for detection of an electromagnetic counterpart of the flare, in an uncertain energy range, because the cascade development may ultimately proceed down to the pair production threshold in the MeV energy range and below. 

Ref.\ \cite{KM3NeT:2025bxl} has analyzed the data on AGN in the vicinity of the neutrino arrival direction. Remarkably, the AGN closest to the source position, TXS 0614-083, has experienced a factor-of-three increase in the X-ray flux, measured by eROSITA during the entire year preceding the ARCA exposure. Fig.\ \ref{fig:sed} shows the spectral energy distribution of the source, with the neutrino flux estimate shown by the black data point. Comparing the upper limits on the source flux in the GeV-TeV energy range from Fermi-LAT telescope (which is two orders of magnitude below the neutrino flux estimate), one can derive a lower bound on either the time delay of the electromagnetic cascade emission, $T_\gamma\gtrsim 100$~yr, or an upper bound on the opening angle of the neutrino emission cone, $\Theta_\nu\lesssim 0.1\Theta_\gamma$, within the scenario in which 100~PeV gamma-rays can escape from the source. In the scenario of an optically thick source, the cascade emission can be emitted in the energy range below 1~MeV, where currently no constraints on the source flux are available (except for the constraints from SVOM, reported in Ref. \cite{KM3NeT:2025bxl}), and it is well possible that a source flare has occurred in this electromagnetic band, but went undetected. 
It is possible that the observed X-ray flare is connected to the neutrino emission but only if the X-rays originated in a wider-angle outflow, and the neutrinos in a collimated beam.

\section{Constraint on a population of flaring neutrino sources}

The combination of \km3net and IceCube data also imposes a constraint on the properties of the population of flaring sources that may produce 100~PeV neutrinos.  We consider a population of flaring sources similar to the source considered in the previous section. For simplicity, we assume that the sources are ``standard candles": each source has the same energy output in flares of identical duration across the entire source population. 

The source at the position of \neutrino is favourably situated in the sky, close to the maximum of the IceCube effective area. This effective area is larger than the sky-average effective area at the reference neutrino energy by a factor-of $\kappa\simeq 3$. If the rate of occurrence of sources with flare flux $F$ and duration $T$, each producing $N_{\rm IC,expected}\simeq \kappa^{-1}20N_{\rm ARCA}\simeq 0.6$ is ${\cal R}$ sources per year, the overall neutrino flux from such source population is $N_{tot}=N_{\rm IC,expected}{\cal R}T_{\rm IC}$ where $T_{\rm IC}\sim 10$~yr is the total IceCube exposure. The non-observation of events with energies in excess of 100~PeV in the ten-year exposure of IceCube imposes a constraint that at most $N_{max}=2.3$ events might have been expected (at the 90\% confidence level) so that the rate of occurrence of neutrinos from the flaring sources on the sky is limited to 
\begin{equation}
    {\cal R}\lesssim \frac{(2.3/0.6)}{T_{\rm IC}}\simeq 0.4\mbox{ yr}^{-1}
\end{equation}

Sources similar to the flaring source that produced the \km3net\ event may be situated at different distances. This leads to a powerlaw type distribution 
\begin{equation}
    \frac{d{\cal R}(N)}{dN}\propto N^{-\alpha}
\end{equation}
so that the number of sources producing flux larger than $F$ (or counts larger than $N_{\rm ARCA}$) is ${\cal R}\propto F^{1-\alpha}\propto N_{\rm ARCA}^{1-\alpha}$. Standard candle sources homogeneously distributed over a cosmological volume have $\alpha=5/2$ down to the minimal event fluence $N_{min}$ corresponding to the sources at the distance $D\sim D_{cosm}$ where $D_{cosm}$ is the cosmological distance scale. The upper bound on the ${\cal R}$ refers to the occurrence rate of sources producing the flux at the level $N\sim N_{min}$. If the source of the ARCA event were nearby, the total fluence from all the source population  would be larger by a factor $\sim D_{cosm}/D$ (where we estimate   $D_{cosm}\sim 20$~Gpc as the luminosity distance up to $z\sim 3$ peak of the AGN and star formation activity) and the upper bound on the flaring source occurrence rate would be tighter:
\begin{equation}
    {\cal R}_{\rm{KM3\mbox{-}230213A\mbox{-}like\mbox{-}flares}}\lesssim 0.4\left[\frac{D}{D_{cosm}}\right]\mbox{ yr}^{-1}
\end{equation}
This upper bound on the flaring events rate can be expressed as a bound on the 
 rate $\rho_s$ of occurrence of sources per unit volume per year. The overall rate of sources within a cosmological distance scale is 
\begin{equation}
    \rho_s=\frac{{
    \cal R}_{\rm{KM3\mbox{-}230213A\mbox{-}like\mbox{-}flares}}}{(4\pi/3)D^3}\lesssim 10^{-5}\left[\frac{D}{D_{cosm}}\right]^{-2}\frac{1}{\mbox{Gpc}^3\mbox{yr}}
\end{equation}
The per-flavour neutrino emissivity of the source population is 
\begin{equation} 
j \lesssim \mathcal{E} \rho_s \approx 4\times 10^{47} \left[\frac{\Theta_\nu}{1^\circ}\right]^2\frac{\mbox{erg}}{\mbox{Gpc}^{3}\mbox{yr}}
\label{eq:emissivity}
\end{equation} 
which corresponds to $j_{\rm iso} \sim 2\times 10^{52}\mbox{erg~Gpc}^{-3}~\mbox{yr}^{-1}$ for isotropically emitting sources. For comparison, the emissivity of high-luminosity GRBs is $j_{\rm iso,HLGRB}~=~3.6~\times~10^{53}~\mbox{erg Gpc}^{-3}~\mbox{yr}^{-1}$~\cite{Dermer:2000yd} and their apparent rate density of $\rho_{\rm HLGRB}\approx  0.8~\mbox{Gpc}^{-3}\mbox{yr}^{-1}$~\cite{Sun:2015bda}.  As a further reference, the local emissivity of Ultra-High-Energy Cosmic Ray (UHECR) sources  is $6\times 10^{53}~\mbox{erg Gpc}^{-3}\mbox{yr}^{-1}$~\cite{PierreAuger:2020kuy}, assuming again isotropic emission. This illustrates the fact that the emissivity of the ARCA neutrino sources is high but not extreme. On the other hand, the sources of UHECRs need to be sufficiently numerous, so that the observed UHECR nuclei, which have a small $\mathcal{O}$(100~Mpc) energy loss length, reach the Earth (the local number density of UHECR sources has been estimated to be higher than $n_{\rm UHECR}\gtrsim 10^4$~Gpc$^{-3}$~\cite{PierreAuger:2013waq}). Therefore, even though the source of the \km3net neutrino definitely accelerates particles to the UHECR energy range, the observed UHECRs cannot be produced by rare sources belonging to the same class as \neutrino.

\section{Discussion}

The combination of the ARCA and IceCube observations is consistent with the possibility of a flaring isolated neutrino source in the direction of the neutrino \neutrino, with overall fluence in the range ${\cal E}\sim 10^{50}-10^{54}$~erg, depending on the assumptions on the anisotropy of the neutrino emission and the distance to the source. This fluence is consistent with a range of transient phenomena, including AGN flares, stellar core collapses that may or may not give rise to GRBs, and tidal disruption events. 
As discussed in Section \ref{sec:electromagnetic}, the neutrino flare might or might not have had an electromagnetic counterpart, depending on the compactness of the source with respect to the process of pair production by gamma rays produced together with the neutrinos. The cascade phenomenon may lead to a different beam size of the electromagnetic emission with respect to that of the neutrinos and to a time delay of this emission that would reduce the electromagnetic flux.

It is well possible that the increased X-ray flux of the AGN nearest to the event position, TXS 0614-083, is the electromagnetic counterpart of the neutrino flare. This is possible if the source is compact. In this case, the cascade development channels the electromagnetic power initially injected in the energy range around 100~PeV can be released in the energy range below the MeV energy of the pair production threshold. Within this interpretation, the X-ray emission must have emerged from a wider angle outflow with opening angle $\Theta_{\gamma} = \sqrt{F_{\nu}/F_{\gamma}} \Theta_{\nu}$, and comparison of the electromagnetic and neutrino flux levels gives 
 a constraint on the anisotropy pattern of the neutrino emission. 
 
Another important aspect of the neutrino signal is that the spectrum of the flaring source is hard and is consistent with originating in a very hard spectrum of protons. The proton spectrum should extend into the domain of Ultra-High-Energy Cosmic Rays (UHECR). Remarkably, modelling of the UHECR flux \cite{PierreAuger:2022atd} suggests that the sources of UHECR should have hard intrinsic spectra, to explain rapid changes in the composition of the flux. If, instead, the proton spectrum isn't particularly hard, the only possibility consistent with the hard neutrino spectrum is that the neutrino was produced in interactions with infrared photons. A natural environment would then be a blazar jet~\cite{Murase:2014foa,Oikonomou:2021akf}.  

A hard proton spectrum in the source may arise from particle acceleration near black holes, as in the model of Ref. \cite{Neronov:2002se} (initially introduced to explain the presence of high-energy electrons in kiloparsec-scale jets of AGN). In this model, the 100~PeV energy gamma-rays produced together with the neutrinos initiate an electromagnetic cascade in the kiloparsec-scale jet. A characteristic feature of the model is that the anisotropy pattern of the electromagnetic and neutrino emission is different because the cascading particles are deflected by magnetic fields present in the jet and/or in the interstellar medium through which the jet propagates \cite{Neronov:2002se}. Neutrino emission within this model has been discussed in Ref.\ \cite{Neronov:2002xv}. A hard proton energy spectrum may also arise from acceleration in shearing flows~\cite{Rieger:2004jz,Rieger:2019uyp,Kimura:2017ubz}. 
These scenarios outline possible origins for the neutrino event KM3-230213A, which -- if the source is transient-- alleviate the tension between the \km3net\ observation and the upper limits of IceCube and the Pierre Auger Observatory.\\

\section*{Acknowledgements}
We thank Martin Lemoine for stimulating discussions. FO thanks the Theory Group at the \emph{Laboratoire Astroparticule et Cosmologie} for the kind hospitality. 

\bibliographystyle{uhecr} 
\bibliography{refs}
\end{document}